\documentclass[12pt]{article}
\usepackage{a4wide}
\usepackage{amssymb}
\usepackage{amsmath}
\usepackage{amstext}
\usepackage{amsfonts}
\usepackage{natbib}

\newcommand {\br}[1]{\left(#1\right)}

\begin{document}

\title{Kernel methods in genomics and computational biology}
\author{Jean-Philippe Vert}
\date{\today}
\maketitle
\abstract{Support vector machines and kernel methods are increasingly popular in genomics and computational biology, due to their good performance in real-world applications and strong modularity that makes them suitable to a wide range of problems, from the classification of tumors to the automatic annotation of proteins. Their ability to work in high dimension, to process non-vectorial data, and the natural framework they provide to integrate heterogeneous data are particularly relevant to various problems arising in computational biology. In this chapter we survey some of the most prominent applications published so far, highlighting the particular developments in kernel methods triggered by problems in biology, and mention a few promising research directions likely to expand in the future.}

\section{INTRODUCTION}
Recent years have witnessed a dramatic evolution in many fields of life science with the apparition and rapid spread of so-called high-throughput technologies, which generate huge amounts of data to characterize various aspects of biological samples or phenomena. To name just a few, DNA sequencing technologies have already provided the whole genome of several hundreds of species, including the human genome \citep{Consortium2001Initial,Venter2001Sequence}; DNA microarrays \citep{Schena1995Quantitative}, that allow the monitoring of the expression level of tens of thousands of transcripts simultaneously, opened the door to functional genomics, the elucidation of the functions of the genes found in the genomes \citep{DeRisi1997Exploring}; recent advances in ionization technology have boosted large-scale capabilities in mass spectrometry, and the rapidly growing field of proteomics, focusing on the systematic, large-scale analysis of proteins \citep{Aebersold2003Mass}. As biology suddenly entered this new era characterized by the relatively cheap and easy generation of huge amounts of data, the urgent need for efficient methods to represent, store, process, analyze, and finally make sense out of these data triggered the parallel development of numerous data analysis algorithms in computational biology. Among them, kernel methods in general, and support vector machines (SVM) in particular, have quickly gained popularity for problems involving the classification and analysis of high-dimensional or complex data. Half a decade after the first pioneering papers \citep{Mukherjee1998Support,Haussler1999Convolution,Jaakkola1999Using}, these methods have been applied to a variety of problems in computational biology, with more than 100 research papers published in 2004 only\footnote{A list of references is available at \texttt{http://cg.ensmp.fr/\~{ }vert/svn/bibli/html/biosvm.html}}. The main reasons behind this fast development, beyond the generally good performances of SVM on real-world problems and ease of use provided by current implementations, involve (i) the particular capability of SVM to resist to high dimensional and noisy data, typically produced by various high-throughput technologies, and (ii) the possibility to process non-vectorial data, such as biological sequences, protein structures or gene networks, and to easily fuse heterogeneous data thanks to the use of kernels. More than a mere application of well-established methods to new datasets, the use of kernel methods in computational biology has been accompanied by new developments to match the specificities and the needs of the field, such as methods for feature selection in combination with the classification of high-dimensional data, the invention of string kernels to process biological sequences, or the development of methods to learn from several kernels simultaneously. In order to illustrate some of the most prominent applications of kernel methods in computational biology and the specific developments they triggered, this chapter focuses on selected applications related to the manipulation of high-dimensional data, the classification of biological sequences, and a few less developed but promising applications. This chapter is therefore not intended to be an exhaustive survey, but rather to illustrate with some examples why and how kernel methods have invaded the field of computational biology so rapidly. The interested reader will find more references in the book by \citet{Schoelkopf2004Kernel} dedicated to the topic. Several kernels for structured data, such as sequences or trees, widely developed and used in computational biology, are also presented in detail in the book by \citet{Shawe-Taylor2004Kernel}.

\section{CLASSIFICATION OF HIGH-DIMENSIONAL DATA}

Several recent technologies, such as DNA microarrays, mass spectrometry or various miniaturized assays, provide thousands of quantitative parameters to characterize biological samples or phenomena. Mathematically speaking, the results of such experiments can be represented by high-dimensional vectors, and many applications involve the supervised classification of such data. Classifying data in high dimension with a limited number of training examples is a challenging task that most statistical procedures have difficulties dealing with, due in particular to the risk of overfitting the training data. The theoretical foundations of SVM and related methods, however, suggest that their use of regularization allows them to better resist to the curse of dimension than other methods. SVM were therefore naturally tested on a variety of datasets involving the classification of high-dimensional data, in particular for the analysis of tumor samples from gene expression data, and novel algorithms were developed in the framework of kernel methods to select a few relevant features for a given high-dimensional classification problem.

\subsection{Tumor Classification from Gene Expression Data}

The early detection of cancer and prediction of cancer types from gene expression data have been among the first applications of kernel methods in computational biology \citep{Mukherjee1998Support,Furey2000Support} and remain prominent. These applications have indeed potentially important impacts on the treatment of cancers, providing clinicians with an objective and possibly highly accurate information to choose the most appropriate form of treatment. In this context, SVM were widely applied and compared with other algorithms for the supervised classification of tumor samples from expression data, of typically several thousands of genes for each tumor. Examples include the discrimination between acute myeloid and acute lymphoblastic leukemia \citep{Mukherjee1998Support}, colon cancer and normal colon tissues \citep{Moler2000Analysis}, normal ovarian, normal non-ovarian and cancer ovarian tissues \citep{Furey2000Support}, melanoma, soft tissue sarcoma and clear cell sarcoma \citep{Segal2003Classification}, different types of soft tissue sarcomas \citep{Segal2003Classificationa}, or normal and gastric tumor tissues \citep{Meireles2003Differentially}, to name just a few. Another typical application is the prediction of the future evolution of a tumor, such as the discrimination between relapsing and nonrelapsing Wilms tumors \citep{Williams2004Prognostic}, the prediction of metastatic or non-metastatic squamous cell carcinoma of the oral cavity \citep{ODonnell2005Gene}, or the discrimination between diffuse large B-cell lymphoma with positive or negative treatment outcome \citep{Shipp2002Diffuse}.

The SVM used in these studies are usually linear hard-margin SVM, or linear soft-margin SVM with a default $C$ parameter value. Concerning the choice of the kernel, several studies observe that nonlinear kernels tend to decrease performance \citep{Ben-Dor2000Tissue,Valentini2002Gene} compared to the simplest linear kernel, which is coherent with the intuition that the complexity of learning non-linear functions in very high dimension does not play in their favor. On the other hand, the choice of hard-margin SVM, sometimes advocated as a default method when data are linearly separable, is certainly worth questioning in more details. Indeed, the theoretical foundations of SVM suggest that in order to learn in high dimension, one should rather increase the importance of regularization as opposed to fitting the data, which corresponds to decreasing the $C$ parameter of the soft-margin formulation. A few recent papers highlight indeed the fact that the choice of $C$ has an important effect on the generalization performance of SVM for classification of gene expression data \citep{Huang2005Gene}. 

A general conclusion of these numerous studies is that SVM generally provide good classification accuracy in spite of the large dimension of the data. For example, in a comparative study of several algorithms for multi-class supervised classification, including naive Bayes, k-nearest neighbors and decision trees, \citet{Li2004comparative} note that ``[SVM] achieve better performance than any other classifiers on almost all the datasets''. However, it is fair to mention that other studies conclude that most algorithms that take into account the problem of large dimension either through regularization, or through feature selection, reach roughly similar accuracy on most datasets \citep{Ben-Dor2000Tissue}. From a practical point of view, the use of the simplest linear kernel and of the soft-margin formulation of SVM seems to be a reasonable default strategy for this application.

\subsection{Feature Selection}
In the classification of microarray data, it is often important, both for classification performance, biomarker identification and interpretation of results, to select only a few discriminative genes among the thousands of candidates available on a typical microarray. While the literature on feature selection is older and goes beyond the field of kernel methods, several interesting developments with kernel methods have been proposed in the recent years, explicitly motivated by the problem of gene selection from microarray data. 

For example, \citet{Su2003RankGene} propose to evaluate the predictive power of a each single gene for a given classification task by the value of the functional minimized by a one-dimensional SVM, trained to classify samples from the expression of only the single gene of interest. This criterion can then be used to rank genes and select only a few with important predictive power. This procedure therefore belongs to the so-called \emph{filter} approach to feature selection, where a criterion (here using SVM) to measure the relevance of each feature is defined, and only relevant features according to this criterion are kept.

A second general strategy for feature selection is the so-called \emph{wrapper} approach, where feature selection alternates with the training of a classifier. The now widely-used recursive feature elimination (RFE) procedure of \citet{Guyon2002Gene}, which iteratively selects smaller and smaller sets of genes and trains SVM, follows this strategy. RFE can only be applied with linear SVM, which is nevertheless not a limitation as long as many features remain, and works as follows. Starting from the full set of genes, a linear SVM is trained and the genes with the smallest weights in the resulting linear discrimination function are eliminated. The procedure is then repeated iteratively starting from the set of remaining genes, and stops when a desired number of genes is reached.

Finally, a third strategy for feature selection, called \emph{embedded} approach, combines the learning of a classifier and the selection of features in a single step. A kernel method following this strategy has been implemented in the joint classifier and feature optimization (JCFO) procedure of \citet{Krishnapuram2004Joint}. JCFO is roughly speaking a variant of SVM with a Bayesian formulation, in which sparseness is obtained both for the features and the classifier expansion in terms of kernel by appropriate choices of prior probabilities. The precise description of the complete procedure to train this algorithm, involving an expectation-maximization (EM) iteration, would go beyond the scope of this chapter and the interested reader is referred to the original publication for further practical details.

Generally speaking, and in spite of these efforts to develop clever algorithms, the effect of feature selection on the classification accuracy of SVM is still debated. Although very good results are sometimes reported, for example for the JCFO procedure \citep{Krishnapuram2004Joint}, several studies conclude that feature selection, for example with procedures like RFE, do not actually improve the accuracy of SVM trained on all genes \citep{Ambroise2002Selection,Ramaswamy2001Multiclass}. The relevance of feature selection algorithms for gene expression data is therefore currently still a research topic, that practitioners should test and assess case by case.

\subsection{Other High-Dimensional Data in Computational Biology}

While early applications of kernel methods to high-dimensional data in genomics and bioinformatics mainly focused on gene expression data, a number of other applications have flourished more recently, some being likely to expand quickly as major applications of machine learning algorithms. For example, studies focusing on tissue classification from data obtained by other technologies, such as methylation assays, to monitor the patterns of cytosine methylation in the upstream regions of genes \citep{Model2001Feature}, or array comparative genomic hybridization (CGH), to measure gene copy number changes in hundreds of genes simultaneously \citep{Aliferis2002Machine} are starting to accumulate. A huge field of application that still barely caught the interest of the machine learning community is \emph{proteomics}, that is, the quantitative study of the protein content of cells and tissues. Technologies such as tandem mass spectromtetry, to monitor the protein content of a biological sample, are now well developed, and classification of tissues from these data is a future potential application of SVM \citep{Wu2003Comparison,Wagner2003Protocols}. Applications in toxicogenomics \citep{Steiner2004Discriminating}, chemogenomics \citep{Bao2002Identifying,Bock2002new} and analysis of single nucleotide polyphormisms \citep{Yoon2003Analysis,Listgarten2004Predictive} are also promising applications for which the capacity of SVM to classify high-dimensional data has only started to be exploited.

\section{SEQUENCE CLASSIFICATION}

The various genome sequencing projects have produced huge amounts of sequence data that need to be analyzed. In particular, the urgent need for methods to automatically process, segment, annotate and classify various sequence data has triggered the fast development of numerous algorithms for strings. In this context, the possibility offered by kernel methods to process any type of data, as soon as a kernel for the data to be processed is available, has been quickly exploited to offer the power of state-of-the-art machine learning algorithms to sequence processing.

Problems that arise in computational biology consist in processing either sets of sequences of a fixed length, or sets of sequences with variable lengths. From a technical point of view the two problems slightly differ: while there are natural ways to encode fixed-length sequences as fixed-length vectors, making them amenable to processing by most learning algorithms, manipulating variable-length sequences is less obvious. In both cases, many successful applications of SVM have been reported, combining ingenious developments of string kernels, sometimes specifically adapted to a given classification task, with the power of SVM. 

\subsection{Kernels for Fixed-Length Sequences}
 Problems involving the classification of fixed-length sequences appear typically when one wants to predict a property along a sequence, such as the local structure or solvent accessibility along a protein sequence. In that case, indeed, a common approach is to use a moving window, that is, to predict the property at each position independently from the others, and to base the prediction only on the nearby sequence contained in a small window around the site of interest. More formally, this requires the construction of predictive models that take a sequence of fixed length as input to predict the property of interest, the length of the sequences being exactly the width of the window.
 
To fix notations, let us denote by $p$ the common length of the sequences, and by $x = x_{1}\ldots x_{p}$ a typical sequence, where each $x_{i}$ is a letter from the alphabet, e.g., an amino-acid. The most natural way to transform such a sequence into a vector of fixed length is to first encode each letter itself into a vector of fixed length $k$, and then to concatenate the codes of the successive letters to obtain a vector of size $pk$ for the whole sequence. A simple code for letters is the following so-called sparse encoding: denoting by $a$ the size of the alphabet, the $i$-th letter of the alphabet is encoded as a vector of dimension $a$ containing only zeros, except for the $i$-th dimension that is set to $1$. For example, in the case of nucleotide sequences with alphabet $\br{A,C,G,T}$, the codes for $A, C, G$ and $T$ would respectively be $(1,0,0,0), (0,1,0,0), (0,0,1,0)$ and $(0,0,0,1)$ and the code for the sequence of length $3$ $AGT$ would be $(1,0,0,0,0,0,1,0,0,0,0,1)$.
Several more evolved codes for single letters have also been proposed. For example, if one has a prior matrix of pairwise similarities between letters, such as widely-used similarity matrices between amino-acids, it is possible to replace the $0/1$ sparse encoding of a given letter by the vector of similarity with other letters; hence the $A$ in the previous example could for instance be represented by the vector $(1,0,0.5,0)$ to emphasize one's belief that $A$ and $G$ share some similarity. This is particularly relevant for biological sequences where mutations of single letters to similar letters are very common. Alternatively, instead of using a prior matrix of similarity, one can automatically align the sequence of interest to similar sequences in a large sequence database, and encode each position by the frequency of each letter in the alignment. As a trivial example, if our previous sequence $AGT$ was found to be aligned to the following sequences: $AGA, AGC, CGT, ATT$, then it could be encoded by the vector $(0.8,0.2,0,0,0,0,0.8,0.2,0.2,0.2,0,0.6)$, corresponding to the respective frequencies of each letter at each position.

In terms of kernel, it is easy to see that the inner product between sparsely encoded sequences is the number of positions with identical letter. In this representation, any linear classifier, such as that learned by a SVM, associates a weight to each feature, that is, to each letter at each position, and the score of a sequence is the sum of the scores of its letters. Such a classifier is usually referred to as a position-specific score matrix in bioinformatics. Similar interpretations can be given for other letter encodings. An interesting extension of these linear kernels for sequences is to raise them to some small power $d$; in that case, the dimension of the feature space used by kernel methods increases, and the new features correspond to all products of $d$ original features. This is particularly appealing for the sparse encoding, because a product of $d$ binary factors is a binary variable equal to $1$ if and only if all factors are $1$, meaning that the features created by the sparse encoding to the power $d$ exactly indicate the simultaneous presence of up to $d$ particular letters at $d$ particular positions. The trick to take a linear kernel to some power is therefore a convenient way to create a classifier for problems that involve the presence of several particular letters at particular positions.
 
A first limitation of these kernels is that they do not contain any information about the order of the letters: they are for example left unchanged if the letters in all sequences are shuffled according to any given permutation. Several attempts to include ordering information have been proposed. For example, \citet{Raetsch2005RASE} replace the local encoding of single letters by a local encoding of $k$ consecutive letters; \citet{Zien2000Engineering} propose an ingenious variant to the polynomial kernel in order to restrict the feature space to products of features at nearby positions only.

A second limitation of these kernels is that the comparison of two sequences only involves the comparison of features at identical positions. This can be problematic in the case of biological sequences, where insertion of deletions of letters are common, resulting in possible shifts within a window. This problem led \citet{Meinicke2004Oligo} to propose a kernel which incorporates a comparison of features at nearby positions, using the following trick: if a feature $f$ (e.g., binary or continuous) appears at position $i$ in the first sequence, and a feature $g$ appears at position $j$ in the second sequence, then the kernel between the two sequences is increased by $K_{0}(f,g) \exp\br{-(i-j)^2/s}$, where $K_{0}(.,.)$ is a basic kernel between the features such as the simple product. When $\sigma$ is chosen very large, then one recovers the classical kernels obtained by comparing only identical positions ($i=j$); the important point here is that for smaller values of $\sigma$, features can contribute positively even though they might be located at different positions on the sequences.

The applications of kernels for fixed-length sequences to solve problems in computational biology are already numerous. For example, they have been widely used to predict local properties along protein sequences using a moving window, such as secondary structure \citep{Hua2001Novel,Guermeur2004kernel}, disulfide bridges involving cysteines \citep{Passerini2004Learning,Chen2004Prediction}, phosphorylation sites \citep{Kim2004Prediction}, interface residues \citep{Yan2004two-stage,Res2005evolution}, or solvent accessibility \citep{Yuan2002Prediction}. Another important field of application is the annotation of DNA, using fixed-length windows centered on a candidate point of interest as an input to a classifier to detect translation initiation sites \citep{Zien2000Engineering,Meinicke2004Oligo}, splice sites \citep{Degroeve2005SpliceMachine,Raetsch2005RASE}, or binding sites of transcription factors \citep{OFlanagan2005Non,Sharan2005motif-based}. The recent interest in short RNA such as antisense oligonucleotides or small interfering RNAs for sequence-specific knockdown of messenger RNAs has also resulted in several works involving classification of such sequences, which have typically a fixed length by nature \citep{Camps-Valls2004Profiled,Teramoto2005Prediction}. Another important application field for these methods is immunoinformatics, including the prediction of peptides that can elicit an immune response \citep{Donnes2002Prediction,Bhasin2004Prediction}, or the classification of immunoglobulins collected from sain or ill patients \citep{Zavaljevski2002Support,Yu2005Classifying}. In most of these applications, SVM lead to comparable if not better prediction accuracy than competing state-of-the-art methods such as neural networks.

\subsection{Kernels for Variable-Length Sequences}

Many problems in computational biology involve sequences of different lengths. For example, the automatic functional or structural annotation of genes found in sequenced genomes requires the processing of amino-acid sequences with no fixed length. Learning from variable-length sequences is a more challenging problem than learning from fixed-length sequences, because there is no natural way to transform a variable-length string into a vector. For kernel methods, this issue boils down to the problem of defining kernels for variable-length strings, a topic that has deserved a lot of attention in the last few years and has given rise to a variety of ingenious solutions summarized in this section.

The most common approach to make a kernel for strings, as for many other types of data, is to design explicitly a set of numerical features that can be extracted from strings, and then to form a kernel as a dot product between the resulting feature vectors. As an example, \citet{Leslie2002spectrum} represent a sequence by the vector of counts of occurrences of all possible $k$-mers in the sequence, for a given integer $k$, effectively resulting in a vector of dimension $a^k$, where $a$ is the size of the alphabet. As an example, the sequence $AACGTCACGAA$ over the alphabet $\br{A,C,G,T}$ is represented by the $16$-dimensional vector $\br{2,2,0,0,1,0,2,0,1,0,0,1,0,1,0,0}$ for $k=2$, where the dimensions are the counts of occurrences of each $2$-mer $AA, AC, ... , TG, TT$ lexicographically ordered. The resulting spectrum kernel between this sequence and the sequence $ACGAAA$, defined as the linear product between the two $16$-dimensional representation vectors, is equal to $9$. It should be noted that although the number of possible $k$-mers easily reaches the order of several thousands as soon as $k$ is equal to $3$ or $4$, classification of sequences by SVM in this high-dimensional space results in fairly good results. A major advantage of the spectrum kernel is its fast computation; indeed, the set of $k$-mers appearing in a given sequence can be indexed in linear time in a trie structure, and the inner product between two vectors is linear with respect to the non-zero coordinates, i.e., at most linear in the total lengths of the sequences. Several variants to the basic spectrum kernel have also been proposed, including for example kernels based on counts of $k$-mers appearing with up to $m$ mismatches in the sequences \citep{Leslie2004Mismatch}.

Another natural approach to vector representation for variable-length strings is to replace each letter by one or several numerical features, such as physico-chemical properties of amino-acids, and then to extract features from the resulting variable-length numerical time series using classical signal processing techniques such as Fourier transforms \citep{Wang2004Weighted-support} or autocorrelation analysis \citep{Zhang2003Classification}. For example, if $h_{1}, \ldots, h_{n}$ denote $n$ numerical features associated to the successive letters of a sequence of length $n$, then the autocorrelation function $r_{j}$ for a given $j>0$ is defined by
$$
r_{j} = \frac{1}{n-j} \sum_{i=1}^{n-j} h_{i} h_{i+j} .
$$
One can them keep a fixed numbers of these coefficients, for example $r_{1}, \ldots , r_{J}$, and create a $J$-dimensional vector to represent each sequence.

A completely different approach for kernel design is to derive them from probabilistic models. Indeed, before the interest on string kernels grew, a number of ingenious probabilistic models had been defined to represent biological sequences or families of sequences, including for example Markov and hidden Markov models for protein sequences, or stochastic context-free grammars for RNA sequences \citep{Durbin1998Biological}. Several authors have therefore explored the possibility to use such models to make kernels, starting with the seminal work of \citet{Jaakkola2000Discriminative} that introduced the \emph{Fisher kernel}. The Fisher kernel is a general method to extract a fixed number of features from any data $x$ for which a parametric probabilistic model $P_{\theta}$ is defined. Here, $\theta$ represents a continuous $d$-dimensional vector of parameters for the probabilistic model, such as transition and emission probabilities for a hidden Markov model, and each $P_{\theta}$ is a probability distribution. Once a particular parameter $\theta_{0}$ is chosen to fit a given set of objects, for example by maximum likelihood, then a $d$-dimensional feature vector for each individual object $x$ can be extracted by taking the gradient in the parameter space of the log-likelihood of the point:
$$
\phi(x) = \nabla_{\theta} \log P_{\theta} (x) .
$$
The intuitive interpretation of this feature vector, usually referred to as the Fisher score in statistics, is that it represents how changes in the $d$ parameters affect the likelihood of the point $x$. In other word, one feature is extracted for each parameter of the model; the particularities of the data point are seen from the eyes of the parameters of the probabilistic model. The Fisher kernel is then obtained as the dot product of these $d$-dimensional vectors, eventually multiplied by the inverse of the Fisher information matrix to render it independent of the parametrization of the model.

A second line of thoughts to make a kernel out of a parametric probabilistic model is to use the concept of covariance kernels \citep{Seeger2002Covariance}, that is, kernels of the form:
$$
K(x,x') = \int P_{\theta}(x) P_{\theta}(x') d\mu(\theta) ,
$$
where $d\mu$ is a prior distribution on the parameter space. Here, the features correspond to the likelihoods of the objects under all distributions of the probabilistic model; objects are considered similar when they have large likelihoods under similar distributions. An important difference with the kernels seen so far is that here, no explicit extraction of finite-dimensional vectors can be performed. Hence for practical applications one must chose probabilistic models that allow the computation of the integral above. This was carried by \citet{Cuturi2005context-tree} who present a family of variable-length Markov models for strings and an algorithm to perform the integral over parameters and models in the same time, resulting in a string kernel with linear complexity in time and memory with respect to the total length of the sequences.

Alternatively, many probabilistic models for biological sequences, such as hidden Markov models, involve a hidden variable that is marginalized over to obtain the probability of a sequence, i.e., can be written as
$$
P(x) = \sum_{h} P(x,h) .
$$
For such distributions, \citet{Tsuda2002Marginalized} introduced the notion of \emph{marginalized kernel}, obtained by marginalizing a kernel for the complete variable over the hidden variable. More precisely, assuming that a kernel for objects of the form $(x,h)$ is defined, the marginalized kernel for observed objects $x$ is given by
$$
K(x,x') = \sum_{h,h'} K\br{(x,h),(x',h')} P(h|x) P(h'|x') .
$$
In order to motivate this definition with a simple example, let us consider a hidden Markov model with two possible hidden states, to model sequences with two possible regimes, such as introns/exons in eukaryotic genes. In that case the hidden variable corresponding to a sequence $x$ of length $n$ is a binary sequence $h$ of length $n$ describing the states along the sequence. For two sequences $x$ and $x'$, if the correct hidden states $h$ and $h'$ were known, such as the correct decomposition into introns and exons, then it would make sense to define a kernel $K\br{(x,h),(x',h')}$ taking into account the specific decomposition of the sequences into two regimes; for example, the kernel for complete data could be a spectrum kernel restricted to the exons, i.e., to positions with a particular state. Because the actual hidden states are not known in practice, the marginalization over the hidden state of this kernel using an adequate probabilistic model can be interpreted as an attempt to apply the kernel for complete data by guessing the hidden variables. As for the covariance kernel, marginalized kernels can often not be expressed as inner products between feature vectors, and require computational tricks to be computed. Several beautiful examples of such kernels for various probabilistic models have been worked out, including hidden Markov models for sequences \citep{Tsuda2002Marginalized,Vert2006Kernels}, stochastic context-free grammars for RNA sequences \citep{Kin2002Marginalized}, or random walk models on graphs for molecular structures \citep{Kashima2004Kernels}.

Following a different line of thought, \citet{Haussler1999Convolution} introduced the concept of \emph{convolution kernels} for objects that can be decomposed into subparts, such as sequences or trees. For example, the concatenation of two strings $x_{1}$ and $x_{2}$ results in another string $x = x_{1}x_{2}$. If two initial string kernels $K_{1}$ and $K_{2}$ are chosen, then a new string kernel is obtained by convolution of the initial kernels following the equation:
$$
K(x,x') = \sum_{x=x_{1}x_{2} , x'=x'_{1}x'_{2}} K_{1}(x_{1},x'_{1})K_{2}(x_{2},x'_{2}) .
$$
Here the sum is over all possible decompositions of $x$ and $x'$ into two concatenated subsequences. The rational behind this approach is that it allows the combination of different kernels adapted to different parts of the sequences, such as introns/exons or gaps/aligned residues in alignment, without knowing the exact segmentation of the sequences. Besides proving that the convolution of two kernels is a valid kernel, \citet{Haussler1999Convolution} gives several examples of convolution kernels relevant for biological sequences; for example, he shows that the joint probability $P(x,x')$ of two sequences under a pair HMM model is a valid kernel, under mild assumptions. This work is extended by \citet{Vert2004Local} where a valid convolution kernel based on the alignment of two sequences is proposed. This kernel, named \emph{local alignment kernel}, is a close relative of the widely used Smith-Waterman local alignment score \citep{Smith1981Identification}, and gives excellent results on the problem of detecting remote homologs of proteins.

Finally, another popular approach to design features and therefore kernels for biological sequences is to ``project'' them onto a fixed dictionary of sequences or motifs, using classical similarity measures, and to use the resulting vector of similarities as feature vector. For example, \citet{Logan2001Study} represent each sequence by a 10,000-dimensional vector indicating the presence of 10,000 motifs of the BLOCKS database; similarly, \citet{Ben-Hur2003Remote} use a vector that indicates the presence or absence of about 500,000 motifs in the eMOTIF database, requiring the use of a trie structure to compute efficiently the kernel without explicitly storing the 500,000 features; and \citet{Liao2003Combining} represent each sequence by a vector of sequence similarities with a fixed set of sequences.

These kernels for variable-length sequences have been widely applied, often in combination with SVM, to various classification tasks in computational biology. Examples including the prediction of protein structural or functional classes from their primary sequence \citep{Ding2001Multi-class,Jaakkola2000Discriminative,Vert2004Local,Karchin2002Classifying,Cai2003Protein}, the prediction of the subcellular localization of proteins \citep{Hua2001Support,Park2003Prediction,Matsuda2005novel}, the classification of transfer RNA \citep{Kin2002Marginalized} and non-coding RNA \citep{Karklin2005Classification}, the prediction of pseudo-exons and alternatively spliced exons \citep{Zhang2003Sequence,Dror2005Accurate}, the separation of mixed plant-pathogen EST collections \citep{Friedel2005Support}, the classification of mammalian viral genomes \citep{Rose2005Correlation}, or the prediction of ribosomal proteins \citep{Lin2002Conserved}.

This short review of kernels developed for the purpose of biological sequence classification, besides highlighting the dynamism of research in kernel methods resulting from practical needs in computational biology, naturally raises the practical question of which kernel to use for a given application. Although no clear answer has emerged yet, some lessons can be learned from early studies. First, there is certainly no kernel universally better than others, and the choice of kernel should depend on the targeted application. Intuitively, a kernel for a classification task is likely to work well if it is based on features relevant to the task; for example, a kernel based on sequence alignments, such as the local alignment kernel, gives excellent results on remote homology detection problems, while a kernel based on the global content of sequences in short subsequences, such as the spectrum kernel, works well for the prediction of subcellular localization. Although some methods for systematic selection and combination of kernels are starting to emerge (see next section), empirical evaluation of different kernels on a given problem seems to be the most common way to chose a kernel. Another important point to notice, besides the classification accuracy obtained with a kernel, is its computational cost. Indeed, practical applications often involve datasets of thousands or tenth of thousands of sequences, and the computational cost of a method can become a critical factor in this context, in particular in an online setting. The kernels presented above differ a lot in their computational cost, ranging from fast linear-time kernels like the spectrum kernel, to slower kernels like the quadratic-time local alignment kernel. The final choice of kernel for a given application often results from a trade-off between classification performance and computational burden.

\section{OTHER APPLICATIONS AND FUTURE TRENDS}

Besides the important applications mentioned in the previous sections, several other attempts to import ideas of kernel methods in computational biology have emerged recently. In this section we highlight three promising directions that are likely to develop quickly in the near future: the engineering of new kernels, the development of methods to handle multiple kernels, and the use of kernel methods for graphs in systems biology.

\subsection{More Kernels}
The power of kernel methods to process virtually any sort of data as soon as a valid kernel is defined has recently been exploited for a variety of data, besides high-dimensional data and sequences. For example, \citet{Vert2002tree} derives a kernel for phylogenetic profiles, that is, a signature indicating the presence or absence of each gene in all sequenced genomes. Several recent works have investigated kernels for protein 3D structures, a topic that is likely to expand quickly with the foreseeable availability of predicted or solved structures for whole genomes \citep{Dobson2003Distinguishing,Borgwardt2005Protein}. For smaller molecules, several kernels based on planar or 3D structures have emerged, with many potential applications in computational chemistry \citep{Kashima2004Kernels,Mahe2005Graph,Swamidass2005Kernels}. This trend to develop more and more kernels, often designed for specific data and applications, is likely to continue in the future because it has proved to be a good approach to obtain efficient algorithms for real-world applications. A nice by-product of these efforts, which is still barely exploited, is the fact that any kernel can be used by any kernel methods, paving the way to a multitude of applications such as clustering \citep{Qin2003Kernel} or data visualization \citep{Komura2005Multidimensional}.

\subsection{Integration of Heterogeneous Data}
Operations on kernels provide simple and powerful tools to integrate heterogeneous data or multiple kernels; this is particularly relevant in computational biology, where biological objects can typically be described by heterogeneous representations, and the availability of a large number of possible kernels for even a single representation raises the question of choice or combination of kernels. Suppose for instance that one wants to perform a functional classification of genes based on their sequences, expression over a series of experiments, evolutionary conservation, and position in an interaction network. A natural approach with kernel methods is to start by defining one or several kernels for each sort of a data, that is, string kernels for the gene sequences, vector kernels to process the expression profiles, etc... The apparent heterogeneity of data types then vanishes as one simply obtains a family of kernel functions $K_{1}, ... , K_{p}$. In order to learn from all data simultaneously, the simplest approach is to define an integrated kernel as the sum of the initial kernels:
$$
K = \sum_{i=1}^p K_{i}.
$$
The rational behind this sum is that if each kernel is a simple dot product, then the sum of dot products is equal to the dot product of the concatenated vectors. In other words, taking a sum of kernels amounts to putting all features of each individual kernel together; if different features in different kernels are relevant for a given problem, then one expects the kernel method trained on the integrated kernel to pick those relevant features. This idea was pioneered by \citet{Pavlidis2002Learning} where gene expression profiles and gene phylogenetic profiles are integrated  to predict the functional classes of genes, effectively integrating evolutionary and transcriptional information. 

An interesting generalization of this approach is to form a convex combination of kernels, of the form:
$$
K = \sum_{i=1}^p w_{i}K_{i} ,
$$
where the $w_{i}$ are nonnegative weights. \citet{Lanckriet2004statistical} propose a general framework, based on semidefinite programming, to optimize the weights and learn a discrimination function for a given classification task simultaneously. Promising empirical results on gene functional classification show that by integrating several kernels, better results than each individual kernel can be obtained.

Finally, other kernel methods can be used to compare and search correlation between heterogeneous data. For example, \citet{Vert2003Extracting} propose to use a kernelized version of canonical correlation analysis (CCA) to compare gene expression data, on the one hand, with the position of genes in the metabolic network, on the other hand. Each type of data is first converted into a kernel for genes, the information about gene positions in the metabolic network being encoded with the so-called diffusion kernel \citep{Kondor2004Diffusion}. These two kernels define embeddings of the set of genes into two Euclidean spaces, in which correlated directions are detected by CCA. It is then shown that the directions detected in the feature space of the diffusion kernel can be interpreted as clusters in the metabolic network, resulting in a method to monitor the expression patterns of metabolic pathways.

\subsection{Kernel Methods in Systems Biology}

Another promising field of research where kernel methods can certainly contribute is \emph{systems biology}, which roughly speaking focuses on the analysis of biological systems of interacting molecules, in particular biological networks.

A first avenue of research is the reconstruction of biological networks from high-throughput data. For example, the prediction of interacting proteins to reconstruct the interaction network can be posed as a binary classification problem -- given a pair of proteins, do they interact or not?--, and can therefore be tackled with SVM as soon as a kernel between \emph{pairs} of proteins is defined. As the primary data available to make the interaction prediction are about each single protein, it is natural to try to derive kernels for pairs of protein from kernel for single proteins. This has been carried out for example by \citet{Bock2001Predicting} who characterize each protein by a vector, and concatenate two such individual vectors to represent a protein pair. Observing that there is usually no order in a protein pair, \citet{Martin2005Predicting} and \citet{Ben-Hur2005Kernel} propose to define a kernel between pairs $(A,B)$ and $(C,D)$ by the equation:
$$
K_{p}\br{(A,B),(C,D)} = K_{i}(A,C)K_{i}(B,D) + K_{i}(A,D)K_{i}(B,C) ,
$$
where $K_{i}$ denotes a kernel for individual protein and $K_{p}$ the resulting kernel for pairs of proteins. The rationale behind this definition is that in order to match the pair $(A,B)$ with the pair $(C,D)$, one can either try to match $A$ with $C$ and $B$ with $D$, or to match $A$ with $D$ and $B$ with $C$.
Reported accuracies on the problem of protein interaction prediction are very high, confirming the potential of kernel methods in this fast-moving field.

A parallel approach to network inference from genomic data has been investigated by \citet{Yamanishi2004Protein}, who show that learning the edges of a network can be carried out by first mapping the vertices, e.g., the genes, onto a Euclidean space, and then connecting the pairs of points which are close to each other in this embedding. The problem then becomes that of learning an optimal embedding of the vertices, a problem known as distance metric learning that recently caught the attention of the machine learning community and for which several kernel methods exist \citep{Vert2005Supervised}.

Finally, several other emerging application in systems biology, such as inference on networks \citep{Tsuda2004Learning} or classification of networks \citep{Middendorf2004Discriminative}, are likely to be subject to increasing attention in the future, due to the growing interest and amount of data related to biological networks.

\section{CONCLUSION}
This brief survey, although far from being complete, highlights the impressive advances in the applications of kernel methods in computational biology in the last 5 years. More than a just importing  well-established algorithms to a new application domain, biology has triggered the development of new algorithms and methods, ranging from the engineering of various kernels to the development of new methods for learning from multiple kernels or for feature selection. The widespread diffusion of easy-to-use SVM softwares, and the ongoing integration of various kernels and kernel methods in major computing environments for bioinformatics, are likely to foster again the use of kernel methods in computational biology, as long as they will provide state-of-the-art methods for practical problems. Many questions remain open, regarding for example the automatic choice and integration of kernels, the possibility to incorporate prior knowledge in kernel methods, and the extension of kernel methods to more general kernels that positive definite, suggesting that theoretical developments are also likely to progress quickly in the near future.

\bibliographystyle{apalike}
\bibliography{bibli}

\end{document}